\documentclass[final,3p,authoryear,11pt]{elsarticle}
\usepackage{amsmath}
\usepackage{amssymb}
\usepackage{endnotes}
\usepackage{lineno}
\usepackage{url}
\usepackage{booktabs}
\usepackage{siunitx} % SI units
\usepackage{xcolor}
\usepackage{todonotes}
\usepackage{ifplatform}

\usepackage{color}
\definecolor{RoyalBlue}{cmyk}{1, 0.50, 0, 0}

\usepackage[colorlinks]{hyperref}
\AtBeginDocument{%
  \hypersetup{
    allbordercolors={1 1 1},
    allcolors={RoyalBlue}}}

\newcommand{\ts}[1]{{#1}\textsuperscript{th}}
\newcommand{\st}[1]{{#1}\textsuperscript{st}}

\journal{ArXiv}

\begin{document}
\begin{frontmatter}

\indent \textcolor{RoyalBlue}{Cite as: Grohmann, C.H., Garcia, G.P.B., Affonso, A.A., Albuquerque, R.W., 2020. Aeolian dune modelling from airborne LiDAR, terrestrial LiDAR and Structure from Motion--Multi View Stereo. \textit{Computers \& Geosciences}. 143:104569. doi:\href{http://dx.doi.org/10.1016/j.cageo.2020.104569}{10.1016/j.cageo.2020.104569}} \\

\vspace{15pt} 

%% Title, authors and addresses
\title{Aeolian dune modelling from airborne LiDAR, terrestrial LiDAR and Structure from Motion--Multi View Stereo}

\author[iee,spam]{Carlos H. Grohmann\corref{cor1}}
\ead{guano@usp.br}
\ead[url]{http://www.iee.usp.br, https://spamlab.github.io}
%\ead[url]{http://carlosgrohmann.com}

\author[igc,spam]{Guilherme P.B. Garcia}
\ead{guilherme.pereira.garcia@usp.br}

\author[iee,spam]{Alynne Almeida Affonso}
\ead{alynne.affonso@usp.br}

\author[iee,spam]{Rafael Walter de Albuquerque}
\ead{rw.albuquerque@gmail.com}

\cortext[cor1]{Corresponding author}

\address[iee]{Institute of Energy and Environment, University of S\~{a}o Paulo (IEE-USP), S\~{a}o Paulo, 05508-010, Brazil}

\address[igc]{Institute of Geosciences, University of S\~{a}o Paulo (IGc-USP), S\~{a}o Paulo, 05508-080, Brazil}

\address[spam]{Spatial Analysis and Modelling Lab (SPAMLab, IEE-USP)}

%% Text of abstract
% Why = The problem/issue that you need to tackle.
% How = Methodology - Methods
% What = Results
% So what = Contribution to knowledge or field.
\begin{abstract}
Sand dunes are commonly regarded as a challenge to traditional photogrammetry due their homogeneous texture and spectral response. In this work we present an evaluation of Structure from Motion--Multi View Stereo (SfM-MVS) to obtain high-resolution elevation data of coastal sand dunes based on images acquired by Remotely Piloted Aircraft (RPA).
A Digital Elevation Model (DEM) of a dunefield in Southern Brazil was generated from 810 photos captured by an RPA at 100~m above the takeoff point in February 2019. Image matching was successful in all areas of the survey due the presence of superficial features (footprints and sandboard tracks) and visibility of the sedimentary stratification, highlighted by heavy minerals. Altimetric accuracy of the SfM-MVS DEM was validated by comparison with Terrestrial LiDAR (TLS) data collected during the same fieldwork campaign of the RPA flights. The SfM-MVS DEM was then compared to an Airborne LiDAR (ALS) DEM from October 2010. 
While the SfM-MVS and TLS DEMs are very similar, without any major difference in elevation or in the reconstruction of topographic features, the SfM-MVS DEM presents a small scale surface roughness not visible in the TLS DEM. The Feature Preserving DEM Smoothing (FPD) algorithm was applied to the SfM-MVS DEM with good results in terms of surface smoothing, but without any significant changes in descriptive statistics and error metrics, with an RMSE of 0.08~m and MAE of 0.06~m for both the original and the FPD-filtered DEM.

Displacement of dune crest lines from the ALS and SfM-MVS DEMs resulted in a migration rate of $\approx$5~m/year between 2010 and 2019, in good agreement with rates derived from satellite images and historical aerial photographs of the same area. Sand volume change in the same period showed a decrease of only 0.2\%, which can be related to the installation of sand fences to promote dune stabilization and sand removal from the front of the dune field to keep a road open to vehicles.
ALS can cover large areas in little time but its high cost still remains a barrier to wider usage, especially by researchers in developing countries. TLS has an intermediate cost but demands more fieldwork and more processing time. In our case we needed three days for the TLS survey and around three weeks to produce a DEM of $\approx$\num{80400}\si{m^2}. On the other hand, we were able to cover $\approx$\num{740900}\si{m^2} with six flight missions in under three hours, with $\approx$13 hours processing time in a medium-range workstation. This makes SfM-MVS a low-cost solution with fast and reliable results for 3D modelling and continuous monitoring of coastal dunes. 
\end{abstract}

\begin{keyword}
Geomorphometry \sep Photogrammetry \sep Digital Elevation Model \sep Point Cloud \sep RPA

\end{keyword}

\end{frontmatter}

% \linenumbers

%% main text
% ----------------------------------------------------------------------------------
% ----------------------------------------------------------------------------------
% ----------------------------------------------------------------------------------
% ----------------------------------------------------------------------------------
\section{Introduction}
\label{sec:intro}

Aeolian dune fields occur in diverse depositional settings, on Earth and on other planetary bodies such as Mars, Venus, Saturn's moon Titan and Pluto \citep{Fryberger1979,Short1988,Wang2002,Livingstone2007,Hayward2007,Radebaugh2008,Bourke2010,Martinho2010,Kreslavsly2017,Hayes2018,Telfer2018}. To better understand these dynamic environments, repeated topographic surveys of the landscape are needed \citep{Conlin2018}. As the sand supply of dune fields is sensitive to patterns of wind and rainfall, changes in dune field volume and morphology can be related to climate change \citep{Gaylord2001,Clemmensen2007,Sawakuchi2008,Tsoar2009,Singhvi2010,Levin2011,Grohmann2013gmorph,Hoover2018}. 

Migration rates of aeolian dunes have been determined with aerial photographs \citep[e.g.,][]{Finkel1961}, orbital imagery \citep{Shrestha2005,Potts2008,Hugenholtz2010,Dong2015,Mendes2015,Mendes2015a,Bhadra2019} or Digital Elevation Models (DEMs\footnote{In this work we use Digital Elevation Model (DEM) in a loose sense to refer to any 3D representation of the land surface, not making a distinction between Digital Terrain Model (DTM) representing the true (bare) ground surface, or Digital Surface Model (DSM) representing a surface that does not necessarily coincide with the ground and may depict man-made structures or vegetation canopy.}) \citep[e.g.,][]{Mitasova2005a}. 

With the growth of Geomorphometry as the practice of terrain modelling and ground-surface quantification \citep{Pike1995,Pike2009,Hengl2008a}, DEMs have became essential tools in landform analysis, as they allow speed, precision and reproducibility to calculation of geomorphometric parameters \citep{Grohmann2004cageo}.

DEMs of aeolian dunes can be constructed by several methods such as traditional field techniques (levelling, Total Station) \citep{Labuz2016}, interpolation of contour lines \citep{Judge2000,Mitasova2005a}, Differential or Real-time kinematic (RTK) GPS points \citep{Mitasova2005a,Pardo-Pascual2005}, LiDAR (Light Detection and Ranging) surveys, either airborne (ALS - Airborne Laser Scanner) \citep{Mitasova2004,Mitasova2005,Mitasova2005a,Vianna2015,Baughman2018}, terrestrial (TLS - Terrestrial Laser Scanner) \citep{Montreuil2013,Feagin2014,Fabbri2017,Sankey2018,Banon2019,Kasprak2019,Lee2019} or mounted on Remotely Piloted Aircrafts (RPAs) \citep{Solazzo2018,Garcin2019}, and Structure from Motion--Multi View Stereo (SfM-MVS) using images collected by handheld cameras, mounted on poles, kites or RPAs \citep{Mancini2013,Goncalves2015,Conlin2018,Duffy2018,Forlani2018,Seymour2018,Solazzo2018,Guisado-Pintado2019,Laporte-Fauret2019,Kasprak2019,Lee2019,ODea2019,Pagan2019,Taddia2019}. A literature review on RPA-based topographic surveys of coastal areas is presented by \cite{Casella2020}. 

In this work we present an evaluation of SfM-MVS to obtain high-resolution elevation data of coastal sand dunes. Altimetric accuracy of the SfM-MVS DEM was validated by comparison with TLS data collected during the same fieldwork campaign of the RPA flights (February 2019). The SfM-MVS DEM was then compared to an ALS DEM from October 2010. The results show almost no change in total volume and a migration rate of $\approx$5~m/year, compatible with those derived from aerial and orbital imagery. 

While 3D modelling of aeolian sand dunes can be a challenge to traditional photogrammetry due to their homogeneous texture and spectral response, the use of SfM-MVS is recommended and the factors that contributed to a successful reconstruction are discussed.

% ----------------------------------------------------------------------------------
% ----------------------------------------------------------------------------------
\subsection{Study area}
\label{sec:study_area}

\begin{figure}[!hbt]
    \centering
    \includegraphics[width=0.95\textwidth]{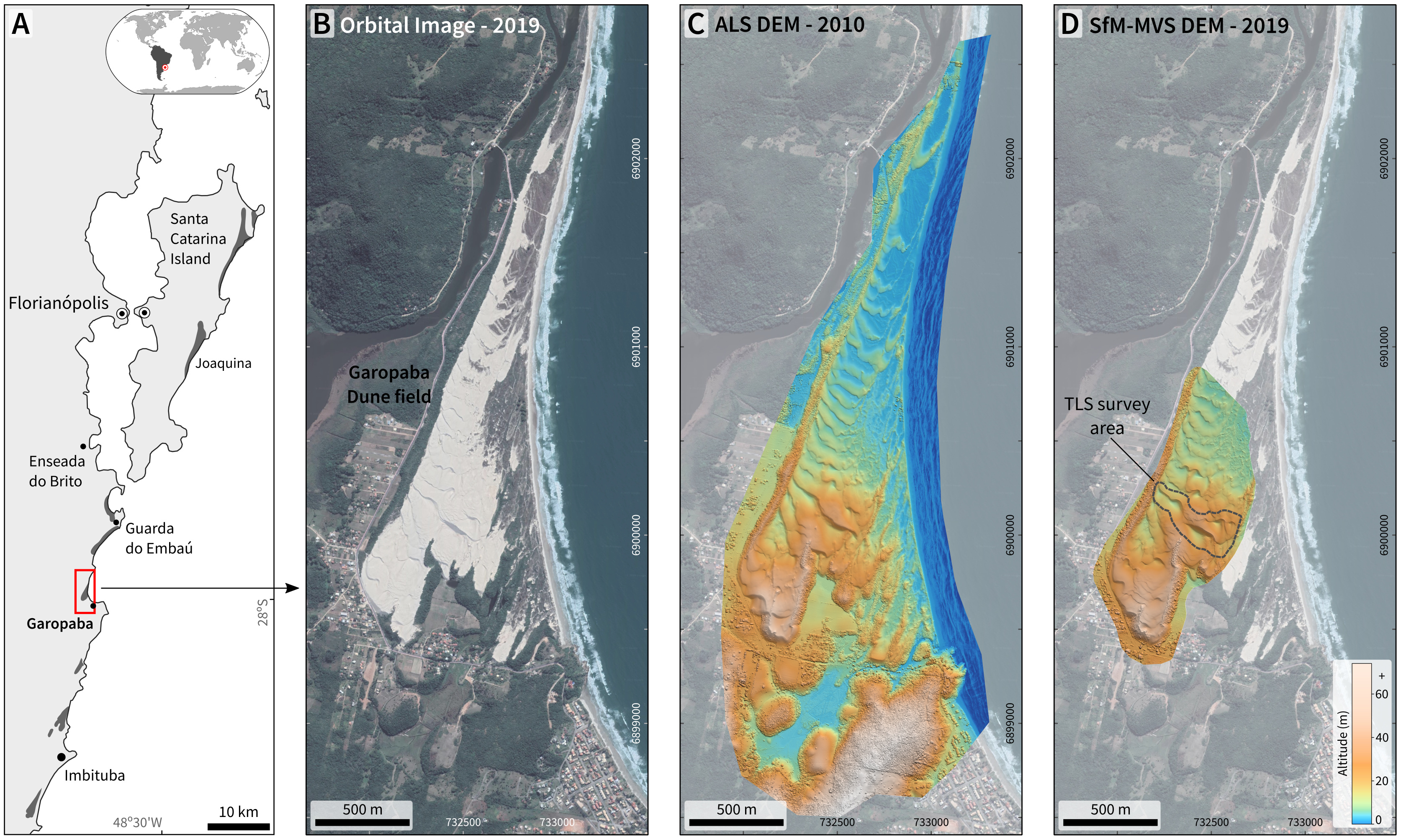}
    \caption{A) Location of study area in southern Brazil; B) Satellite image of the Garopaba dune field (image date: 07-30-2017); C) Digital Surface Model produced by ALS (2010); D) DEM produced by SfM-MVS (2019), with TLS survey area shown. Elevation colour scale is the same for C and D. Shaded relief illumination: N\ang{025}, \ang{30} above horizon. Dune field satellite imagery \textcopyright 2019 Maxar Technologies, powered by Google. Coordinate system for B/C/D and following figures: UTM zone 22 southern hemisphere, WGS84 datum.}
    \label{fig:loc_dems}
\end{figure}

The study area, located in Santa Catarina State, southern Brazil (Fig.~\ref{fig:loc_dems}A-B), comprises barrier-lagoon depositional systems with associated dune fields \citep{Angulo2006,Giannini2007} which evolved during the Late Holocene as a result of wind strength intensification and sand supply increase in southern Brazilian coast \citep{Mendes2015,Mendes2015a}. The Garopaba (or Siri\'u) dune field is composed of unvegetated and vegetated aeolian dunes. The unvegetated dunes are represented by mostly barchanoid chains, while the vegetated ones include parabolic dunes, blowouts and foredunes \citep{Martinho2006,Hesp2007}. 

There are significant differences of wind field along the southern Brazilian coast; while the dominant and prevailing direction is from the S at Joaquina (located $\approx$\num{45}\si{km} north of Garopaba in Santa Catarina Island), it is from the NE at Farol de Santa Marta, $\approx$\num{70}\si{km} south of the study area \citep{Dillenburg2006,Hesp2007,Truccolo2011,Mendes2015}. At Garopaba, winds from the North are responsible for dune migration \citep{Mendes2015,Mendes2015a}.

% \textit{(Figure \ref{fig:loc_dems} about here)}\\

% fig - location and DEMs
% \begin{figure}[pos=h]
% \clearpage

% ----------------------------------------------------------------------------------
% ----------------------------------------------------------------------------------
% ----------------------------------------------------------------------------------
% ----------------------------------------------------------------------------------
\section{Methods}
\label{sec:methods}

This section presents the datasets, methods and tools used in this study. A flowchart of the analysis steps is in the Supplemental Material. Table \ref{tbl:pcs_stats} shows, for each kind of data used in this paper (ALS, SfM-MVS, TLS), area of the interpolated DEM, number of points and density of points within that area.

% Table 1 - point clouds
\begin{table}%[!hbt]
    \caption{Overview of datasets used in this study. See text for details and Supplemental Material for maps of point density.}
    \label{tbl:pcs_stats}
    \begin{center}
    \resizebox{0.65\textwidth}{!} {
    \begin{tabular}{lrrr}
    \toprule
    Data                            & DEM Area (\si{m^2}) & \# points        & points/\si{m^2} \\
    \midrule
    ALS (full)                      &       \num{4434722} &   \num{11574555} &      \num{2.6}  \\
    ALS (SfM area, \st{1} returns)  &        \num{740922} &    \num{2380005} &      \num{3.2}  \\
    SfM-MVS (full)                  &        \num{740922} &  \num{344769434} &    \num{465.3}  \\
    SfM-MVS (thin \ts{125} pt)      &        \num{740922} &    \num{2378399} &      \num{3.2}  \\
    SfM-MVS (TLS area)              &         \num{80413} &   \num{28158102} &    \num{350.1}  \\
    SfM-MVS (10~cm grid)            &         \num{80413} &    \num{8079569} &    \num{100.5}  \\
    TLS (full)                      &         \num{80413} & \num{1187708492} &  \num{14770.1}  \\
    TLS (2~cm filter)               &         \num{80413} &  \num{170141709} &   \num{2115.8}  \\
    TLS (10~cm grid)                &         \num{80413} &    \num{7039501} &     \num{87.5}  \\
    \bottomrule
    \end{tabular}
    } %end resizebox
    \end{center}
\end{table}
% \clearpage

\subsection{Airborne LiDAR}
\label{sec:methd_als}
Airborne LiDAR (ALS) data were collected on October 2010 by Geoid Laser Mapping Co. using an Optech ALTM 3100 sensor with a saw-tooth scanning pattern, density of about one point per 0.5 m$^2$, measured from an altitude of $\approx$\num{1200}~m ($\approx$\num{4000}~ft). Raw LiDAR data (with up to four laser pulses) were processed by Geoid and delivered with vertical accuracy of 0.15~m ($1\sigma$) and horizontal accuracy of 0.5~m ($1\sigma$).

ALS data (LiDAR \st{1} returns) were imported into GRASS-GIS \citep{Neteler2012} as vector points and interpolated with bilinear splines \citep{Brovelli2004,Brovelli2004a} to create a DEM with 0.5~m spatial resolution (Fig.~\ref{fig:loc_dems}C).

% ----------------------------------------------------------------------------------
% ----------------------------------------------------------------------------------
\subsection{Fieldwork and Ground Control Points}
\label{sec:methd_gcp}
\begin{figure}[!hbt]
    \centering
    \includegraphics[width=0.95\textwidth]{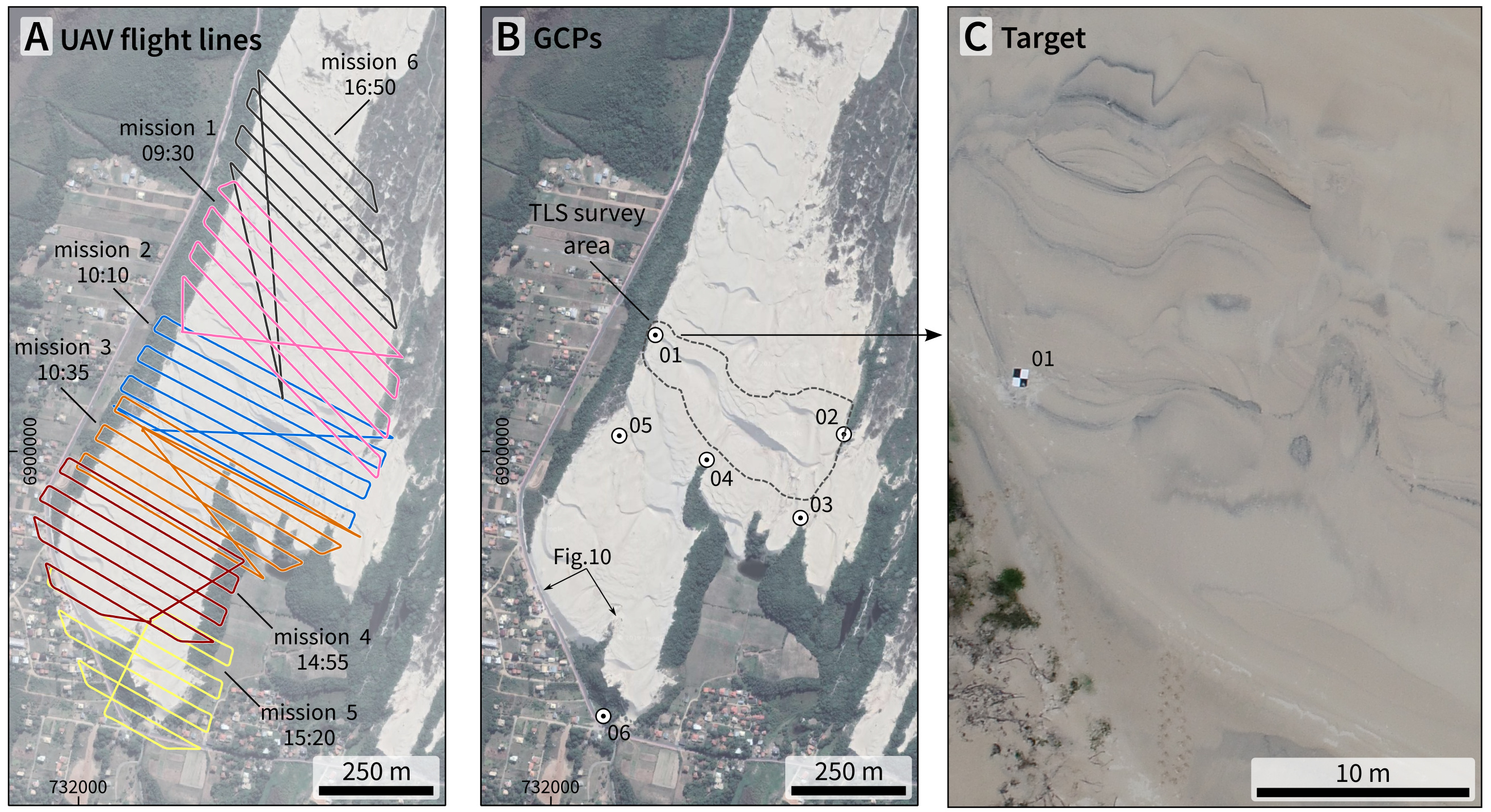}
    \caption{A) Flight missions executed over the dune field; B) Ground Control Points used to georeference the SfM-MVS outputs and TLS data; C) Target (GCP 01), sedimentary structures and superficial features seen in the RPA photo.}
    \label{fig:flights_targets}
\end{figure}

Fieldwork for TLS and SfM-MVS surveys was conducted on February 2019. Six targets were deployed within the dune field area (Fig.~\ref{fig:flights_targets}B) and their coordinates were determined by Differential Global Positioning System (DGPS), to serve as Ground Control Points (GCPs) for georeferencing the SfM-MVS outputs and the TLS point cloud \citep{Harwin2015}.

Each target measured $\approx$80x60~cm in a black and white chequered pattern and was clearly visible in the photos (Fig.~\ref{fig:flights_targets}C). A Spectra Precision SP60 DGPS was used in a base-rover static configuration and raw data was post-processed in Survey Office\endnote{\url{https://spectrageospatial.com/survey-office}} 4.10 software, using the Imbituba Station of the Brazilian GPS Network as reference. The processing reports are available in the Supplemental Material.

% fig - flight missions, GCPs, target photo DJI_0204_RGB_250dpi.jpg
% \begin{figure}[pos=h]
% \clearpage

% ----------------------------------------------------------------------------------
% ----------------------------------------------------------------------------------
\subsection{Terrestrial LiDAR}
\label{sec:methd_tls}
Terrestrial LiDAR (TLS) data were collected with a FARO\texttrademark\ Laser Scanner Focus\textsuperscript{3D} S120, a geodetic laser scanner with distance measurement based on phase shift of infrared light (905~nm), maximum range of 120~m, and ranging error of $\pm$2~mm at 10~m distance at 90\% reflectivity \citep{FARO2013}. The scanner was set at resolution of ``1/5'' and quality of ``3x'', resulting in a point spacing of 7.67~mm at a distance of 10~m, scan time of two minutes and 28.4 million points per scan (this model does not acquire images). Five spherical targets provided with the equipment were arranged on the ground at $\approx$10~m from the TLS and re-positioned in a `leapfrog' scheme during the survey, so that each consecutive scan was able to capture at least two spheres from the previous one. In three days of field work, 110 scans were collected, covering an area of $\approx$\num{80400}~m$^2$ (Fig.~\ref{fig:loc_dems}D).

TLS data were processed in FARO Scene 7.1\endnote{\url{https://knowledge.faro.com/Software/FARO_SCENE/SCENE}}. Each scan was registered to its adjacent ones manually using the spherical targets as references. Georreferencing of the point cloud was based on two DGPS points located at the extremities of the surveyed area (Fig.~\ref{fig:flights_targets}B). Referencing with only two points was possible because this TLS model has an integrated dual axis compensator to automatically level the captured scan data, so the control points were used for an affine transformation (translation/scale/rotation) in 2D space. 

To overcome the heterogeneous distribution of data common to terrestrial LiDAR, with a very high density of points near the scanner, the full point cloud was subsampled in FARO Scene with a minimum distance filter of 2~cm between points. To eliminate duplicate points and compensate for small differences in the alignment of individual scans, this point cloud was gridded to a raster in GRASS-GIS using the mean elevation value of LiDAR points within 10~cm cells (\texttt{r.in.xyz} module). To fill empty (null) cells, the raster was converted to vector and a DEM with 10~cm spatial resolution was created by interpolation with bilinear splines (Fig.~\ref{fig:tls_sfm_shades}A).

% ----------------------------------------------------------------------------------
% ----------------------------------------------------------------------------------
\subsection{SfM-MVS}
\label{sec:methd_sfm}

Images for the SfM-MVS reconstruction were acquired by a DJI Phantom 4 Pro RPA. The aircraft digital camera has an 1'' CMOS 20MP sensor, global shutter, \ang{84} FOV and 8.8~mm focal distance (24~mm at 35~mm equivalent). Images can be saved as JPEG or RAW, \num{5472}$\times$\num{3648}~px (3:2 ratio). Flight missions were planned and executed using the MapPilot app\endnote{\url{https://support.dronesmadeeasy.com}} with height above takeoff point of 100~m (image footprint 150$\times$100~m, pixel size $\approx$2.7~cm) and 75\% overlap along and across-track.

Six missions were flown, covering an area of $\approx$\num{869000}~m$^2$ with 810 images. The camera angle was set to \ang{-80} (i.e., \ang{10} from nadir). Figure~\ref{fig:flights_targets}A shows flight paths and starting time for each mission (UTC-3). Weather conditions during fieldwork were of dark skies with light rains scattered throughout the day.

The SfM-MVS workflow \citep[e.g., ][]{Westoby2012,Viana2018bjgeo,James2019} was processed in Agisoft Metashape Pro version 1.5.1\endnote{\url{https://www.agisoft.com}}. In the SfM step, images were aligned with `High' accuracy. To avoid doming effects in the reconstructed surface \citep[e.g.,][]{James2014}, camera alignment optimization was performed considering a marker accuracy of 0.005~m, following Agisoft's recommendations\endnote{\url{https://www.agisoft.com/index.php?id=31}}. The MVS reconstruction was set to `High' quality and `aggressive' depth filtering. The processing report is available in the Supplemental Material.

For the altimetric comparison with TLS data, the full SfM-MVS point cloud was imported into GRASS-GIS in the same manner of the TLS data: gridded by the mean elevation in 10~cm cells, converted to vector and interpolated with bilinear splines to a DEM with 10~cm resolution (Fig.~\ref{fig:tls_sfm_shades}B).

For the dune migration and volume analysis, the full SfM-MVS point cloud was subsampled (thinned) with LAStools \citep{Isenburg2019} by extracting every 125\textsuperscript{th} point, imported into GRASS-GIS as vector points and interpolated with bilinear splines to a DEM with 0.5~m resolution (Fig.~\ref{fig:loc_dems}D). The thinning value was determined after experimentation, and the goal was to obtain a similar number of points, within the interpolation area, for the ALS and SfM point clouds (Table~\ref{tbl:pcs_stats}).

% ----------------------------------------------------------------------------------
% ----------------------------------------------------------------------------------
\subsubsection{Accuracy of SfM-MVS DEM}
\label{sec:methd_accur}

The vertical accuracy of a DEM can be computed from the differences between the dataset being analyzed and co-located values from an independent source of higher accuracy \citep{Willmott2005,Wechsler2007,Hebeler2009,Reuter2009,Baade2016}. To evaluate the accuracy of the SfM-MVS reconstruction, the TLS DEM was considered as the reference.

Mean Absolute Error (MAE) and Root Mean Square Error (RMSE) are metrics that been widely used in the Geosciences to measure the accuracy of DEMs \citep[e.g., ][]{Nikolakopoulos2006,Willmott2006,Smith2015,Gesch2016,Satge2016,Grohmann2013gmorph,Grohmann2018rse}. MAE (Eq.~\ref{eq:mae}) and RMSE (Eq.~\ref{eq:rmse}) were calculated from all pixels within a mask designed to avoid areas with vegetation or without TLS data.

\begin{equation}
\label{eq:mae}
MAE = \dfrac{1}{n}\textstyle\sum_{i=1}^{i=n}(|z_{TLS} - z_{SfM}|)
\end{equation}

\begin{equation}
\label{eq:rmse}
RMSE = \sqrt{\dfrac{1}{n}\textstyle\sum_{i=1}^{i=n}\Big[(z_{TLS} - z_{SfM})^{2}\Big]}
\end{equation}

% \vskip 0.5cm

% ----------------------------------------------------------------------------------
% ----------------------------------------------------------------------------------
\subsubsection{Surface roughness and DEM smoothing}
\label{sec:methd_denoise}

Surface roughness characterizes elevation variations over a particular scale \citep{Grohmann2010ieee,Berti2013,Grohmann2014encyclopedia,Smith2014}. In this paper surface roughness was calculated as the standard deviation of slope in a moving-window filter, as it provide good results in identifying terrain features and is not sensitive to spurious data \citep{Grohmann2010ieee}. 

Low-pass filters are usually applied to DEMs to remove or reduce roughness \citep{Reuter2009,Gallant2011,Lindsay2019}, but sharp edges such as dune crests will be modified as well \citep{Barash2002,Grohmann2009cageo}. To retain the sharpness of breaks-of-slope in the filtered DEM edge-preserving (or de-noise) procedures \citep[e.g.,][]{Sun2007,Stevenson2009,Lindsay2019} must be employed. 

The effect of de-noising the SfM-MVS DEM was evaluated by applying the Feature Preserving DEM Smoothing (FPD) algorithm \citep{Lindsay2019}, with different parameter settings, to a sub-area of the DEM (Fig.~\ref{fig:sfm_tls_rough}) and evaluating the change of RMSE from the TLS DEM and of Circular Variance of Aspect \citep[CVA --][]{Lindsay2019} for each set of parameters. CVA measures the variability of aspect, or surface shape complexity, within a neighborhood; its value is 0.0 in smooth areas approaching 1.0 in areas of complex topography (i.e., high surface roughness). The three parameters to be set in the FPD method are the filter kernel size ($k$), the normal difference threshold angle ($t$), and the number of elevation-update iterations ($i$) \citep[for a detailed explanation of the algorithm and parameters' definitions, see][]{Lindsay2019}. 

The tests were done by changing one parameter while keeping the other two fixed, and then calculating CVA for each DEM with filter sizes of $3\times3$ up to $41\times41$ pixels. 

In the first experiment, $t$ ranged from \ang{5} to \ang{45} (in \ang{5} increments) with $k=15$ and $i=5$. Next, $k$ changed from $5\times5$ up to $51\times51$ with $t=\ang{20}$ and $i=5$. Last, $i$ varied between 3 and 30 iterations whith $k=15$ and $t=\ang{20}$.

After the sub-area tests, $k$, $t$ and $i$ values were selected and FPD was applied to the entire SfM-MVS DEM. Error metrics (RMSE, MAE) of the original and smoothed SfM-MVS DEM were calculated considering the TLS DEM as reference.

% ----------------------------------------------------------------------------------
% ----------------------------------------------------------------------------------
\subsection{Dune Migration and Sand volume}
\label{sec:methd_migra}

Dune migration can be evaluated from multi-temporal data such as aerial photographs \citep{Finkel1961,Stafford1971,Mendes2015,Baughman2018}, satellite images \citep{Hoover2018,Dong2015,Yang2019} or LiDAR DEMs \citep{Mitasova2004,Mitasova2005,Mitasova2005a,Baughman2018}. Dune migration between the 2010 (ALS) and 2019 (SfM-MVS) surveys was determined as the displacement of dune crest lines.

\begin{figure}[!hbt]
    \centering
    \includegraphics[width=0.95\textwidth]{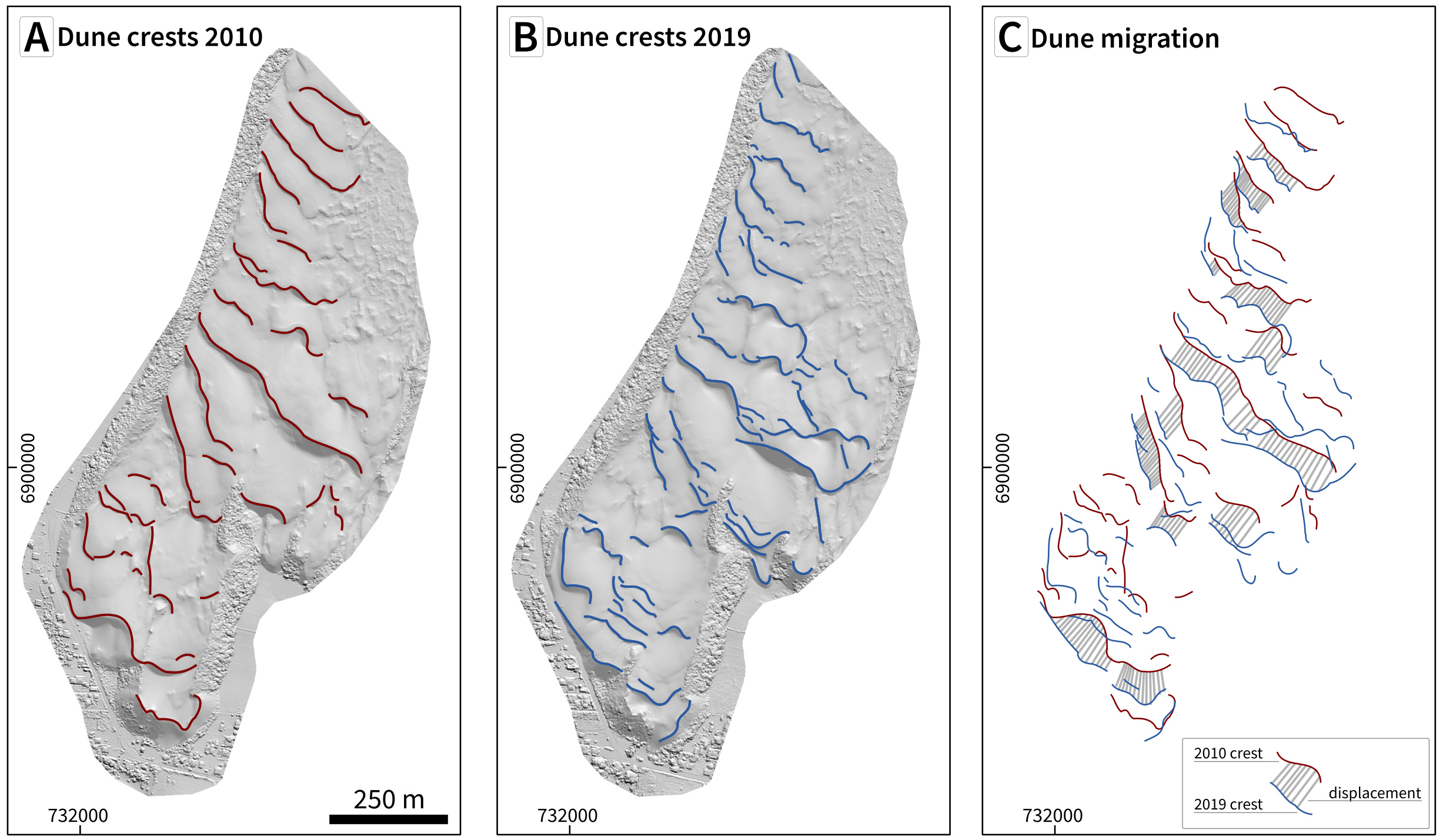}
    \caption{Determination of dune migration between 2010 and 2019 surveys. A) dune crests of 2010, over shaded relief image of ALS DEM; B) dune crests of 2019, over shaded relief image of SfM-MVS DEM; C) displacement lines (grey) connecting crest lines.}
    \label{fig:dune_migration}
\end{figure}

For each survey, surface roughness of the DEM was calculated as the standard deviation of slope \citep{Grohmann2010ieee} in a 5x5 pixels neighbourhood (2.5$\times$2.5~m); crest lines were drawnn in QGIS version 3.8 \citep{QGIS} following the high-roughness crests (see Supplemental Material); lines connecting the crests were draw approximately parallel to the S-SW migration direction \citep{Hesp2007,Mendes2015} (Fig.~\ref{fig:dune_migration}) and saved in shapefile format. Azimuth and length of each displacement line were calculated with Python version 3.7.4 \citep{Python3} using the \texttt{ogr} module of the GDAL library \citep{GDAL} to access vector geometries. Mean azimuth was calculated according to \cite{Fisher1993}.

Sand volume was calculated with the GRASS-GIS \texttt{r.volume} module \citep{Hinthorne1988}. This module calculates volume by summing cell values within a given area and then multiplying by the area occupied by those cells. An elevation of 0~m (zero) was used as a reference base level.

% fig - dune migration
% \begin{figure}[pos=h]

% ----------------------------------------------------------------------------------
% ----------------------------------------------------------------------------------
\subsection{Data Analysis}
\label{sec:methd_data}

In order to streamline the process and ensure reproducibility \citep{Barnes2010}, data analysis was performed in GRASS-GIS version 7.6 \citep{Neteler2012,GRASS2019} through Jupyter notebooks \citep{Kluyver2016,Rule2018a} using the Pygrass library \citep{Zambelli2013} to access GRASS' datasets. The FPD algorithm is implemented in the open-source geospatial analysis platform WhiteboxTools \citep{Lindsay2017}. Statistical analyses were performed with the Python libraries Rasterio, Scipy, Numpy, Pandas, Seaborn and Matplotlib \citep{Oliphant2006,Hunter2007,McKinney2011,Gillies2019,SciPy2013,seaborn2016}.

% ----------------------------------------------------------------------------------
% ----------------------------------------------------------------------------------
% ----------------------------------------------------------------------------------
\section{Results}
\label{sec:results}

% -----------------
\subsection{TLS and SfM-MVS}
\label{sec:results_sfm_tls}

The SfM-MVS survey resulted in data with 2.77~cm resolution (1.1~px reprojection error) and RMSE of 0.7~cm in longitude and 0.4~cm in latitude, and 0.5~cm in elevation for the residuals of control points (see Supplemental Material). The TLS point clouds were combined into a single dataset with a registration error of 4.9~mm. 

\begin{figure}[!hbt]
    \centering
    \includegraphics[width=0.95\textwidth]{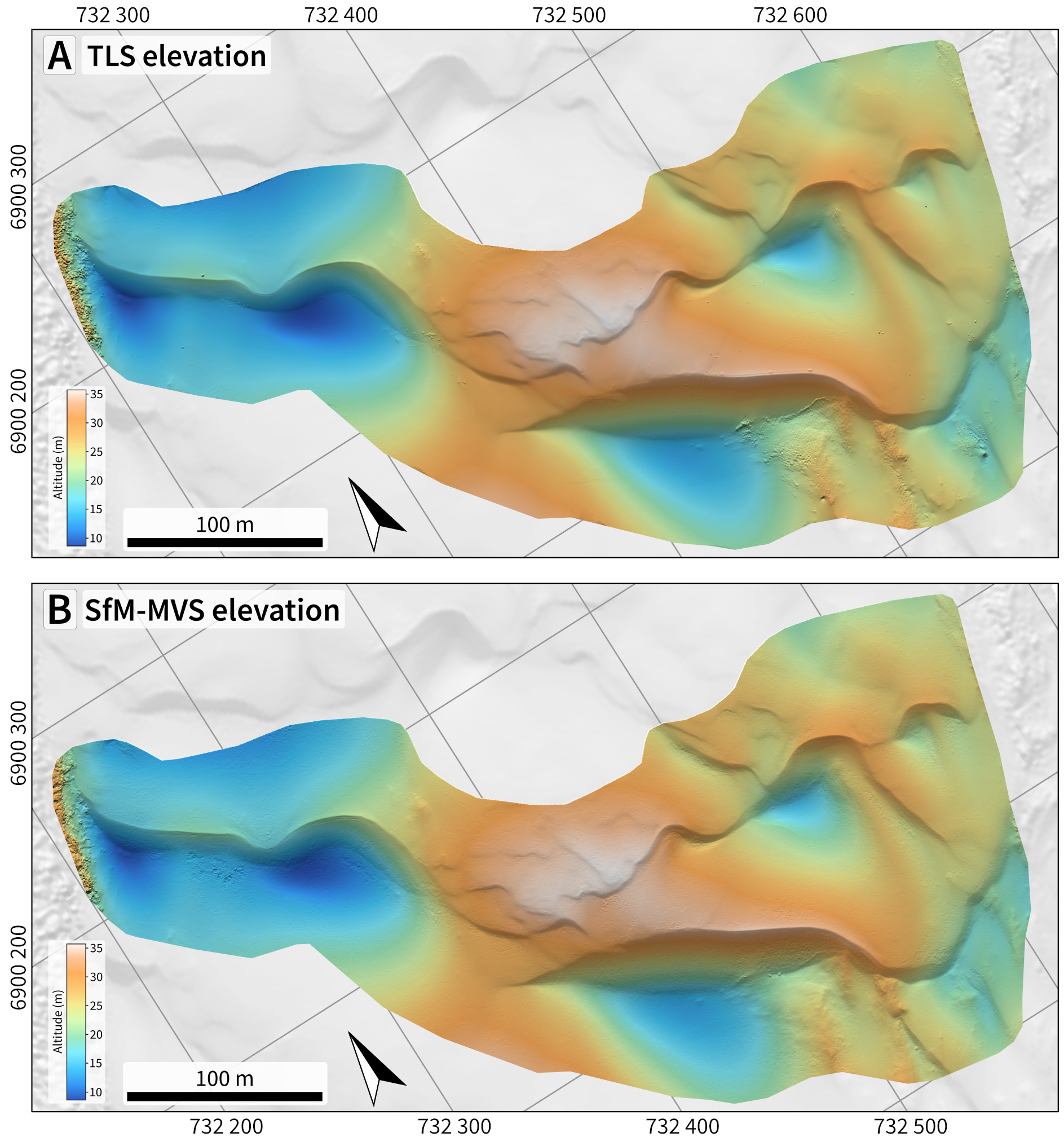}
    \caption{A) TLS DEM; B) SfM-MVS DEM. Elevation colour scale is the same for A and B. Shaded relief illumination: N\ang{025}, \ang{30} above horizon.}
    \label{fig:tls_sfm_shades}
\end{figure}

The DEMs produced from the TLS and SfM-MVS data are presented in Fig.~\ref{fig:tls_sfm_shades}. The surfaces are very similar, without any major difference in elevation or in the reconstruction of topographic features. Upon a closer inspection, the SfM-MVS DEM presents a small scale surface roughness not visible in the TLS DEM. To visually evaluate this difference, surface roughness of the DEMs was calculated as the standard deviation of slope in a 5x5 pixels neighbourhood (0.5$\times$0.5~m). 

% fig - TLS + SfM DEMs
% \begin{figure}[pos=h]
% \clearpage

\begin{figure}[!hbt]
    \centering
    \includegraphics[width=0.95\textwidth]{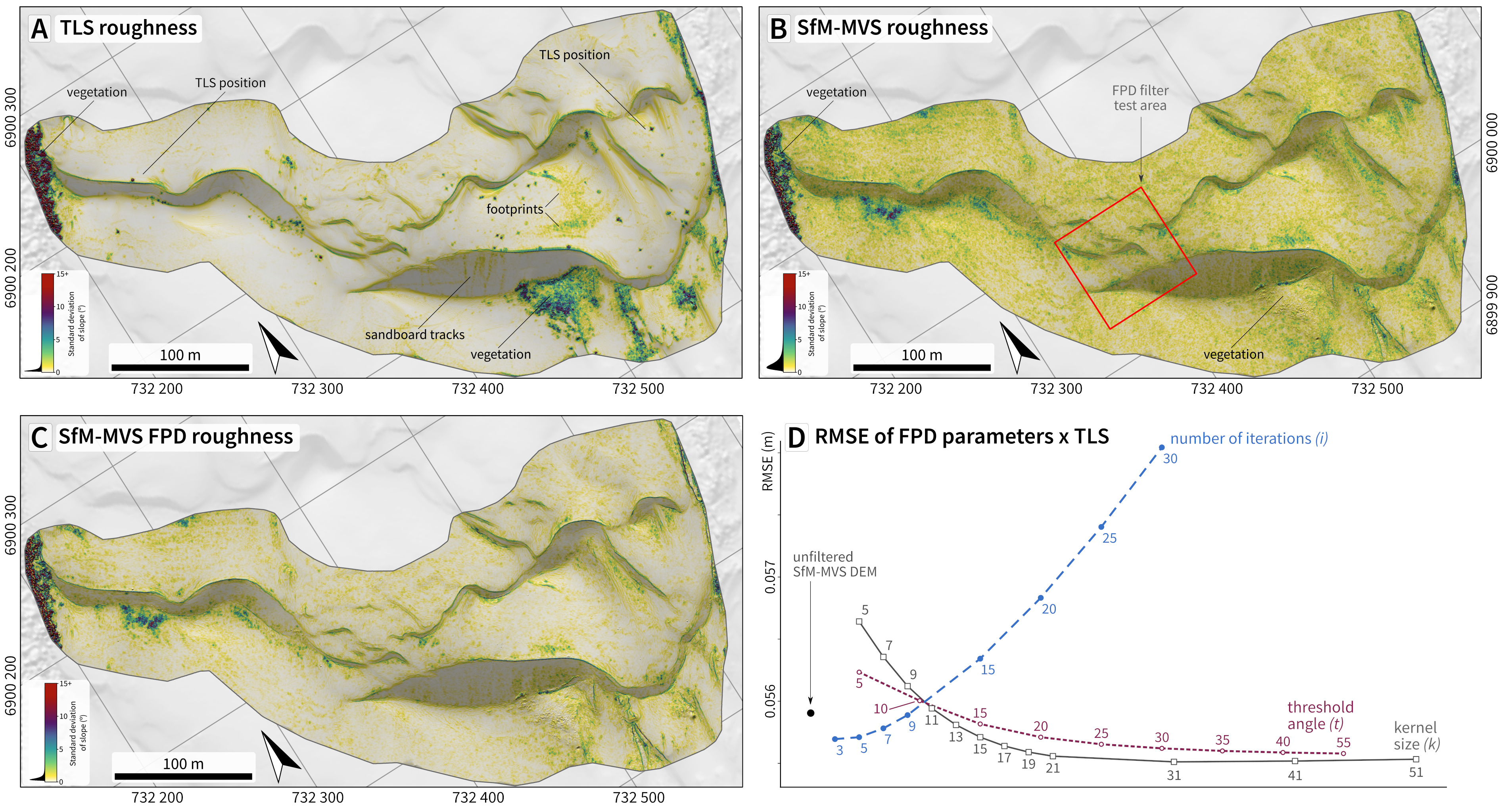}
    \caption{Surface roughness maps, calculated as the standard deviation of slope in a 5x5 window: A) TLS; B) SfM-MVS; C) SfM-MVS after FPD filter ($k=17$, $t=\ang{20}$, $i=5$); D) RMSE plot of each FPD DEM with TLS DEM according to FPD parameters. Roughness colour scale is the same for A, B and C.}
    \label{fig:sfm_tls_rough}
\end{figure}

The TLS DEM has a smooth surface, with higher roughness values on vegetated areas and over some of the places where the TLS equipment was positioned (Fig.~\ref{fig:sfm_tls_rough}A). These spots can be related to a small mismatch between adjacent scans, where in one there is no data (under the scanner), so the gridding procedure cannot compensate the difference and the result is a small circular patch of the terrain slightly above or below its surroundings. Dune crests are well marked by above-average roughness. Footprints and track marks are also visible, with lower roughness values. 

The SfM-MVS DEM shows a widespread distribution of low and average roughness values (Fig.~\ref{fig:sfm_tls_rough}B). While the dune crests can be identified, track marks are no longer visible and the patch of vegetation near the sandboard tracks cannot be discriminated based on its roughness. A set of footprints seen in the central-eastern portion of the TLS roughness map is not visible in the SfM-MVS roughness because the SfM-MVS survey was carried out before the TLS survey could cover that area. We see this roughness as a noise inherent to the SfM process due to small errors in geolocation as well as to the consumer-grade quality of the photographic camera \citep{Mosbrucker2016}.

% fig - roughness
% \begin{figure}[pos=h]

Applying the FPD algorithm to a sub-area of the SfM-MVS DEM shows the RMSE between the smoothed DEM and the TLS DEM decreasing with larger FPD kernel sizes and higher angular treshold values, but increasing with the number of interactions (Fig.~\ref{fig:sfm_tls_rough}C). Considering these results, we applied FPD to the SfM-MVS DEM with $k=17\times17$ $t=\ang{20}$ and $i=5$ (Fig.~\ref{fig:sfm_tls_rough}D). An evaluation of Circular Variance of Aspect for each parameter leads to similar conclusions, and is presented in the Supplemental Material. 

\begin{figure}[!hbt]
    \centering
    \includegraphics[width=0.95\textwidth]{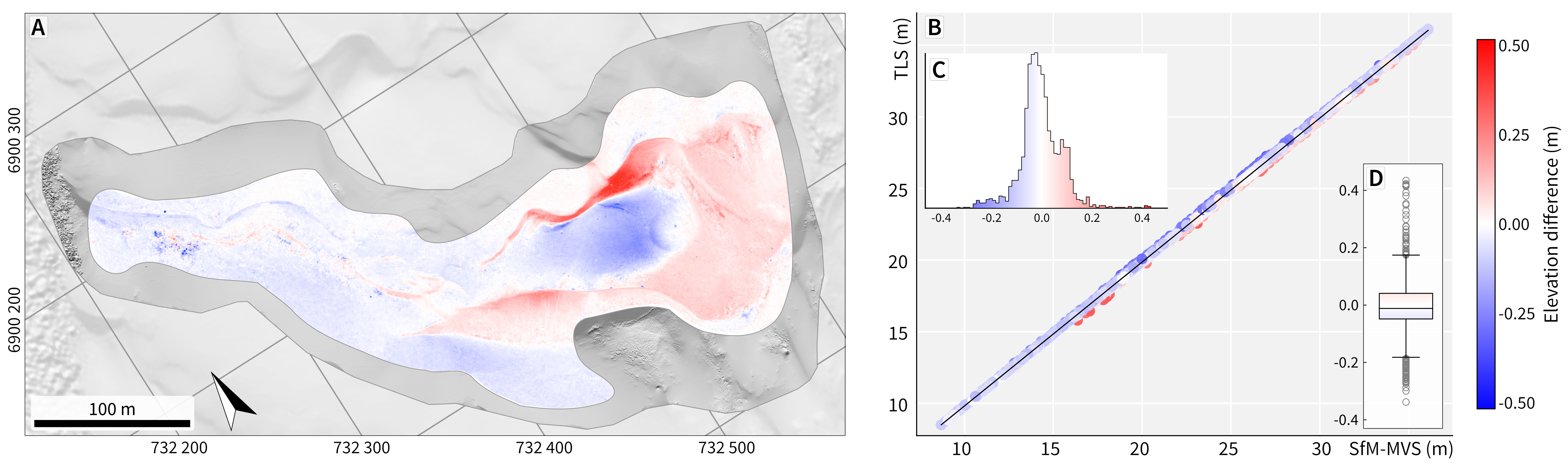}
    \caption{Differences between TLS and SfM-MVS: A) Map of difference values (all pixels); B) Scatterplot; C) Histogram (bins=60); D) Boxplot. Data for B,C,D from a set of \num{2000} random pixels.}
    \label{fig:sfm_tls_hists}
\end{figure}

The vertical accuracy of the SfM-MVS DEM, calculated for all pixels within the mask shown in Fig.~\ref{fig:sfm_tls_hists}A resulted in RMSE of 0.08~m and MAE of 0.06~m for both the original and the FPD-filtered DEM. 

Descriptive statistics of the TLS and SfM-MVS DEMs are very similar (Table~\ref{tbl:stats_tls_sfm}). Considering all pixels of the DEMs, elevation differences range from -1.5~m to +0.5~m, with mean of 0.0~m and standard deviation of 0.08~m; with a random sample of \num{2000} pixels, elevation differences range from -0.3~m to +0.5~m, with mean of 0.0~m and standard deviation of 0.08~m, The negative differences below -0.5~m can be disregarded as they represent a small fraction of the total (312 pixels out of $\approx\num{4.8e6}$ pixels). 

A scatterplot of elevations (\num{2000} pixels, TLS $\times$ SfM-MVS, Fig.~\ref{fig:sfm_tls_hists}B) shows minimal dispersion of points, with an R$^2$ of 0.999 (see Supplemental Material). The histogram of differences (Fig.~\ref{fig:sfm_tls_hists}C) has a bimodal distribution, with $\approx$55\% of the values below zero, indicating that, in general, the SfM-MVS DEM has higher elevations than the TLS DEM, and the boxplot of differences (Fig.~\ref{fig:sfm_tls_hists}D) shows 106 points classified as outliers (values beyond $\pm0.17$~m).

% Table 2 - stats TLS x SfM
\begin{table}%[!hbt]
    \caption{Descriptive statistics of the TLS, SfM-MVS DEMs and of differences. Elevation units in metres (m).}
    \label{tbl:stats_tls_sfm}
    \begin{center}
     \resizebox{0.65\textwidth}{!} {
        \begin{tabular}{lrrrrrrrrr}
        \toprule
          {} &  min &   max &  mean &  median &  std.dev. &  skewness &  kurtosis &   25\%quant. &   75\%quant. \\
        \midrule
all pixels \\
        \midrule
           TLS &  8.490 & 36.172 & 23.937 & 24.753 &  6.519 & -0.277 & -0.750 & 18.708 & 28.712 \\
       SfM-MVS &  8.440 & 36.153 & 23.938 & 24.774 &  6.531 & -0.283 & -0.754 & 18.691 & 28.713 \\
   SfM-MVS FPD &  8.445 & 36.152 & 23.938 & 24.774 &  6.532 & -0.283 & -0.754 & 18.691 & 28.713 \\
 Diff. SfM TLS & -1.506 &  0.519 &  0.001 & -0.008 &  0.085 &  0.701 &  4.801 & -0.045 &  0.046 \\
 Diff. FPD TLS & -1.510 &  0.509 &  0.001 & -0.008 &  0.085 &  0.703 &  4.849 & -0.045 &  0.046 \\
         \midrule
\num{2000} pixels \\
        \midrule
           TLS &  8.526 & 36.023 & 24.015 & 24.840 &  6.455 & -0.277 & -0.725 & 18.863 & 28.639 \\
       SfM-MVS &  8.483 & 35.990 & 24.010 & 24.830 &  6.468 & -0.281 & -0.730 & 18.822 & 28.654 \\
   SfM-MVS FPD &  8.477 & 35.989 & 24.010 & 24.826 &  6.468 & -0.281 & -0.729 & 18.824 & 28.654 \\
 Diff. SfM TLS & -0.338 &  0.434 & -0.004 & -0.012 &  0.084 &  0.344 &  3.172 & -0.048 &  0.041 \\
 Diff. FPD TLS & -0.344 &  0.429 & -0.004 & -0.012 &  0.084 &  0.344 &  3.186 & -0.048 &  0.042 \\
        \bottomrule
        \end{tabular}
     } % end resizebox
    \end{center}
\end{table}

% fig - histograms TLS x SfM
% \begin{figure}[pos=h]

\subsection{ALS and SfM-MVS}
\label{sec:results_als-sfm}
\begin{figure}[!hbt]
    \centering
    \includegraphics[width=0.95\textwidth]{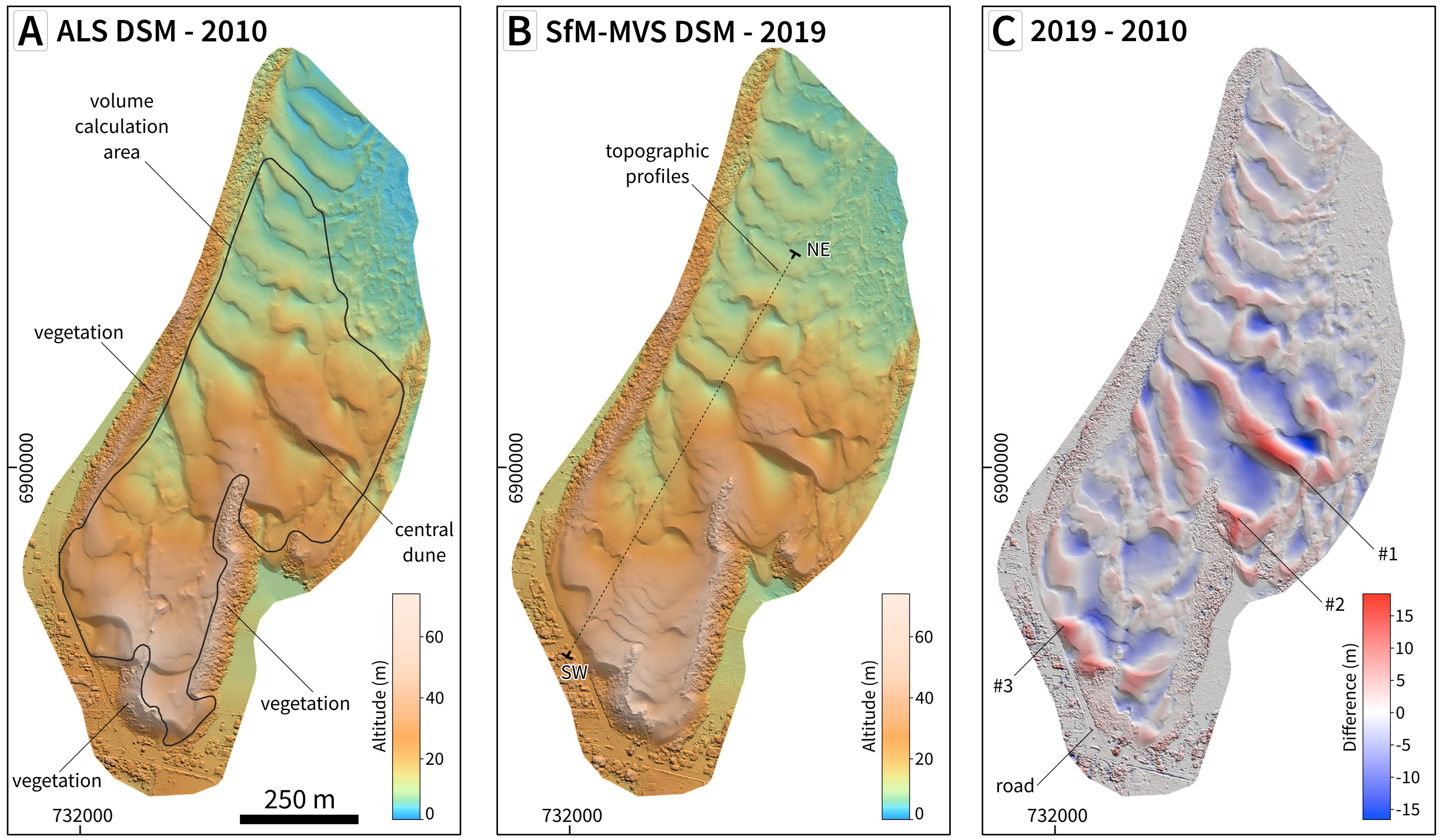}
    \caption{A) ALS DEM (2010), with volume calculation area polygon; B) SfM-MVS DEM (2019), with topographic profiles location; C) DEM of differences (2019-2010). Numbers in C are discussed in the text.}
    \label{fig:als_sfm_dod}
\end{figure}

Besides a good correlation to the TLS DEM, the full SfM-MVS DEM (Fig.~\ref{fig:als_sfm_dod}B) shows a good fit with elements of the landscape that didn't experienced significant change between the surveys, such as the road bordering the dune field to west and southwest (in grey in Fig.~\ref{fig:als_sfm_dod}C, indicating no elevation difference).

Comparison of the 2010 ALS and 2019 SfM-MVS DEMs was carried out based on: 1) descriptive statistics of the DEMs; 2) differences between the DEMs; 3) sand volume within an area and 4) displacement of dune crests.

Differences between the DEMs were calculated by subtracting the elevations of the ALS DcrestEM from the SfM-MVS DEM. Positive values represent areas where the SfM surface has higher elevations than the ALS one, and vice-versa.

The ALS and SfM-MVS DEMs (0.5~m resolution) and the differences between the two surfaces, are presented in Fig.~\ref{fig:als_sfm_dod}. In the studied area, dunes are mainly barchanoids with lee side towards south west. Elevation reaches its highest ($\approx$58~m) in the southern portion, likely due the influence of an underlying palaeotopography \citep{Giannini2007}.

% fig - ALS DEM, SfM DEM, DoD
% \begin{figure}[pos=h]

\begin{figure}[!hbt]
    \centering
    \includegraphics[width=0.95\textwidth]{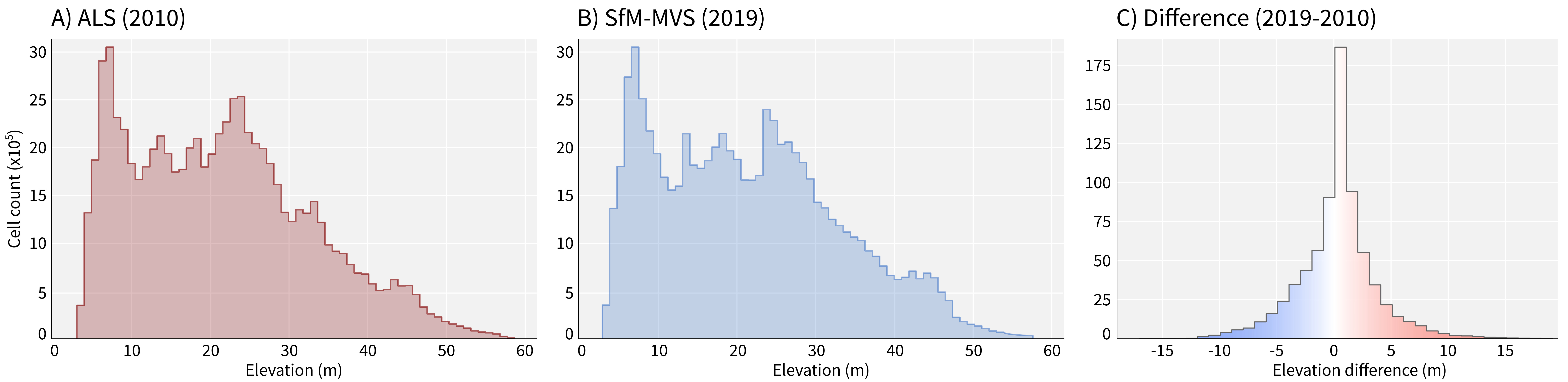}
    \caption{Histograms: A) ALS DEM (bins=60); SfM-MVS DEM (bins=60); C) DEM of differences (bins=40).}
    \label{fig:als_sfm_dod_hists}
\end{figure}

\begin{figure}[!hbt]
    \centering
    \includegraphics[width=0.95\textwidth]{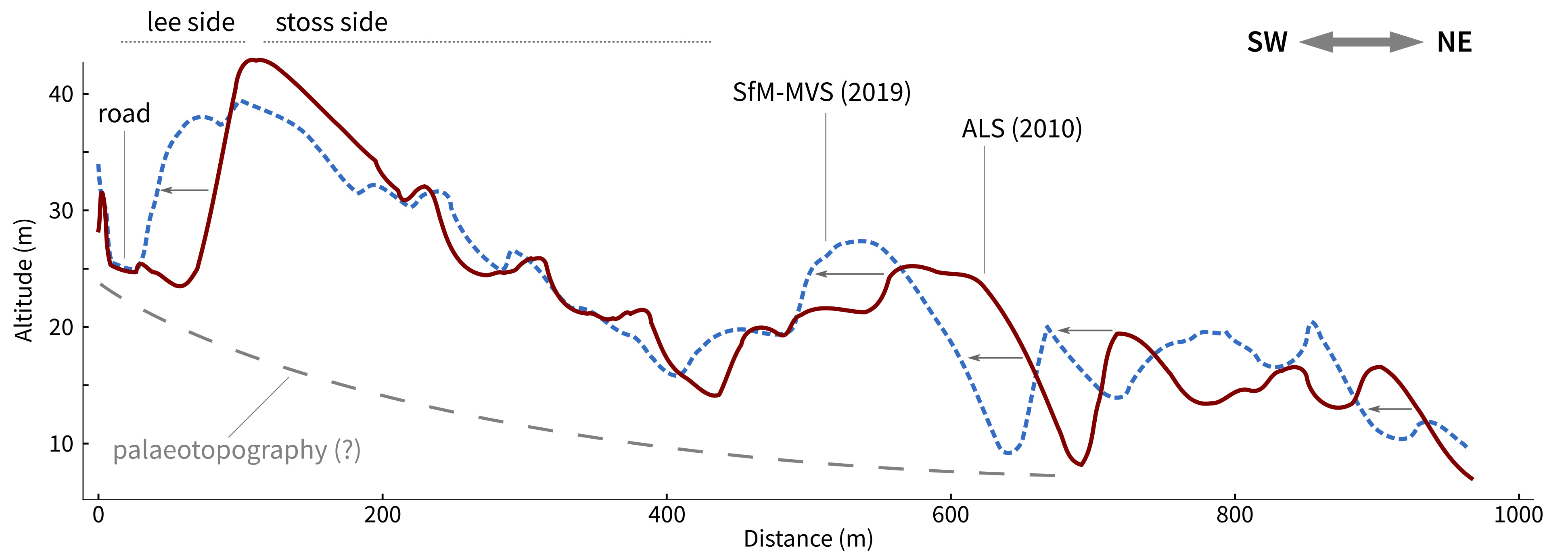}
    \caption{Topographic profiles across the dune field (location in Fig.~\ref{fig:als_sfm_dod}B.)}
    \label{fig:profiles}
\end{figure}

% stats - table, histograms, DoD
Descriptive statistics are presented in Table~\ref{tbl:dems_stats} and histograms of elevation values in Fig.~\ref{fig:als_sfm_dod_hists}. The DEM of differences between 2019 and 2010 DEMs is in Fig.~\ref{fig:als_sfm_dod}C; positive values are in red and negative values in blue. Topographic profiles (location in Fig.~\ref{fig:als_sfm_dod}B) are in Fig.~\ref{fig:profiles}.

The DEMs have similar values of maximum, standard deviation, skewness, kurtosis and quantiles. The SfM-MVS DEM shows slightly higher mean and minimum values. Elevation differences between the DEMs range from -16.95~m to +23.15~m, with mean and median of $\approx$0.0~ms. Some notable differences are indicated as \#1,\#2 and \#3 in Fig.~\ref{fig:als_sfm_dod}C: \#1 marks the highest positive difference (where the SfM-MVS surface is above the ALS), related to the migration of a large `central dune' with accumulation of sand towards a vegetated ridge in \#2; \#3 shows the migration of the dune field over the road. In this place, the town hall needs to remove the sand periodically to keep the road open.

% Table 3 - descriptive stats
\begin{table}%[!hbt]
    \caption{Descriptive statistics of the ALS, SfM-MVS DEMs and of differences between the two surfaces.}
    \label{tbl:dems_stats}
    \begin{center}
    \resizebox{0.75\textwidth}{!} {
    \begin{tabular}{rrrrrrrrrrr}
    \toprule
    {} &  min &   max &  mean &  median &  std.dev. &  skewness &  kurtosis &   25\%quant. &   75\%quant. \\
    \midrule
    ALS     &   2.69 & 58.88 & 21.34 &  20.65 &  11.59 & 0.51 & -0.41 & 11.64 & 28.77 \\
    SfM-MVS &   2.89 & 58.63 & 21.64 &  20.67 &  11.66 & 0.45 & -0.57 & 11.63 & 29.40 \\
    Diff.   & -16.95 & 23.15 &  0.31 &   0.39 &   3.48 & 0.16 &  2.77 & -1.26 &  1.80 \\
    \bottomrule
    \end{tabular}
    } %end resizebox
    \end{center}
\end{table}
% \clearpage

% fig - histograms
% \begin{figure}[pos=h]

% volume
The polygon for volume calculation encloses only unvegetated areas in both surveys (see Fig.~\ref{fig:als_sfm_dod}A). Using the ALS and SfM DEMs with 0.5~m resolution, the calculated sand volumes were \num{9035115.45}~\si{\cubic\metre} for 2010 and \num{9010844.95}~\si{\cubic\metre} for 2019 (a decrease of \num{24270.50}~\si{\cubic\metre} or 0.2\%).

% displacement
Dune crest displacement lines drawn over the DEMs (see Fig.~\ref{fig:dune_migration}) yielded a mean azimuth of \ang{215.5} and mean length of $\approx$44.5~m (mean: 44.3~m, median: 44.7~m, see Supplemental Material for statistical analysis of azimuth and length).

A mean length of 44.5~m in 9 years corresponds to a dune migration rate of $\approx$5~m/year. We consider these rates to be in agreement with rates of 6-7~m/year from \cite{Mendes2015} and \cite{Mendes2015a}, which were derived from interpretation of historical aerial photographs and satellite images with coarser spatial resolution.

% topo profiles
Topographic profiles (Fig.~\ref{fig:profiles}) illustrate dune movement from 2010 to 2019, with migration of the lee side and relatively less change over the stoss side of large compound dunes.

% fig - profiles
% \begin{figure}[pos=h]
% \clearpage

% ----------------------------------------------------------------------------------
% ----------------------------------------------------------------------------------
% ----------------------------------------------------------------------------------
% ----------------------------------------------------------------------------------
\section{Discussions and Conclusions}
\label{sec:discussion}

In this work we presented an evaluation of SfM-MVS in high-resolution topographic modelling of coastal sand dunes.

Although sand dunes are commonly regarded as a challenge to traditional photogrammetry due their homogeneous texture and spectral response, yielding poor results in image matching \citep{Baltsavias1999}, recent literature on close-range photogrammetry/SfM-MVS of coastal areas report good results in surface reconstruction \citep{Goncalves2015,Goncalves2018,Duffy2018,Laporte-Fauret2019,Puijenbroek2017,Guisado-Pintado2019,Pitman2019}.

In this research, image matching was successful in all areas of the survey due the presence of superficial features (footprints and sandboard tracks) and visibility of the sedimentary stratification, highlighted by heavy minerals (Fig.~\ref{fig:flights_targets}C). 

One factor that positively influenced the RPA survey was the weather. A cloudy sky provided a diffuse illumination, without `hard' shadows, and the scattered light rain ensured that the sand was humid, without the presence of a layer of loose sand over the dunes, which would mask the stratifications and other features in the photos \citep{Guisado-Pintado2019}.

We believe that the lack of texture in aerial photographs and satellite images is more related to ground resolution (i.e., pixel size) than the spectral or morphological characteristics of aeolian dunes, as a pixel area of one square metre can be enough to `average-out' small textural features and prevent good image matching. This is an issue to be seen in the context of the everlasting matter of scale in remote sensing and geomorphometry: pixel size \textit{vs.} spatial structure (size) of landforms \citep[e.g.,][]{Woodcock1987,Wood1996a,Gallant1997,Goodchild1997,Marceau1999,Hengl2006,Kamal2014}. Large continental dunes, for instance, have been successfully modelled with 30~m-resolution images from Landsat and ASTER \citep{Levin2004,Bullard2011}.

To validate the use of an SfM-MVS DEM, a TLS DEM was used as reference for altimetric accuracy. The comparison resulted in RMSE of 0.08~m and MAE of 0.06~m. The TLS DEM has a smooth appearance, with well-marked dune crests and vegetated areas, while the SfM-MVS DEM shows a small-scale roughness that hinders visual identification of small features such as footprints. 

Although it does not influence the comparison with ALS data, this roughness can be an issue if the objective of the research is the classification of landforms based on geomorphometric parameters, such as the identification of dune crests based on surface curvature \citep{Mitasova2005a,Hardin2014}. The FPD de-noise algorithm \citep{Lindsay2019} was applied to the SfM-MVS DEM with good results in terms of surface smoothing, without any significant changes in descriptive statistics and error metrics.

Dune crests interpreted from the ALS DEM were compared to crests from the SfM-MVS DEM and resulted in a migration rate of $\approx$5~m/year, in good agreement with rates derived from satellite images and historical aerial photographs of the same area \citep{Mendes2015,Mendes2015a}. 

\begin{figure}[!hbt]
    \centering
    \includegraphics[width=0.95\textwidth]{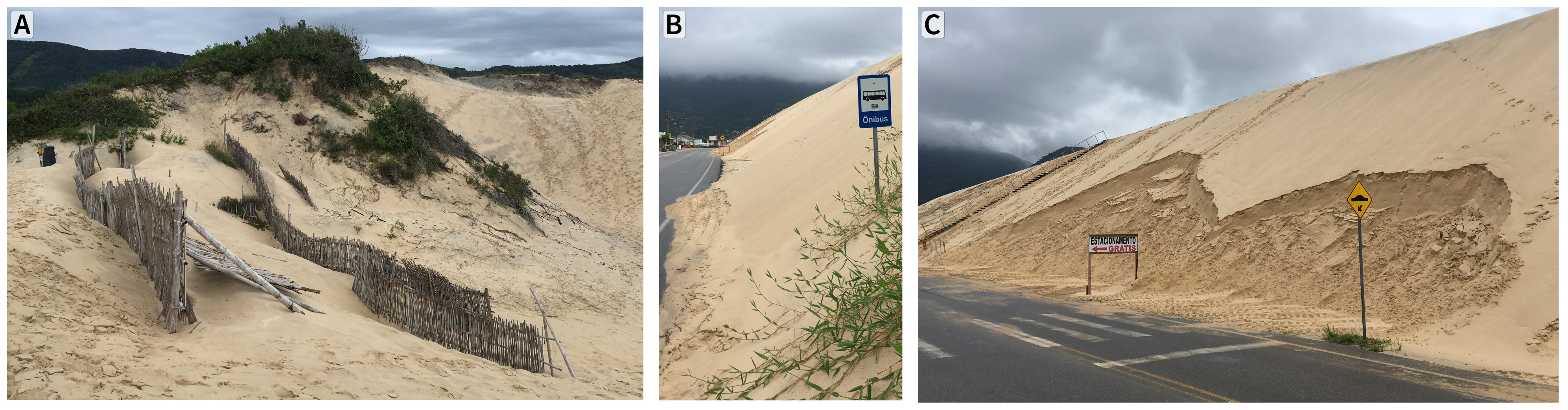}
    \caption{A) Sand fences installed to promote dune stabilization; B, C) Front of the dune field showing migration over road and signs. Location of photos on Fig.~\ref{fig:flights_targets}B.}
    \label{fig:fences}
\end{figure}

Volumes calculated from the ALS and SfM-MVS DEMs show a difference of 0.2\% between 2010 and 2019. Such small variation is within reported uncertainties for SfM-MVS reconstructions \citep{Draeyer2014,Rhodes2017,Gupta2018} and may be related to the the installation of sand fences to promote dune stabilization and the constant removal of sand from the road in front of the dune field (Fig.~\ref{fig:fences}). Future studies can explore the spatial distribution of differences in the DEMs to evaluate the sedimentary budget (\textit{influx$-$efflux}) of the dune field \citep{Sankey2018a,Sankey2018,Kasprak2019}.

% fig - sand fences and migration over road
% \begin{figure}[pos=h]

When comparing these different approaches to aeolian dune surface modelling (ALS, TLS and SfM-MVS) we must consider not only the accuracy of final products (DEMs), but also the time required to acquire the data and process it to a GIS-ready format. 

ALS might be acquired in little time, but it is by far the most expensive, imposing a serious constrain on repeated surveys, especially for researchers in developing countries or without access to state-funded coastal monitoring programs. 

TLS has an intermediate cost of acquisition (since the equipment can be rented and operated by the research team) but it demands more fieldwork and more processing time. In our case we needed three days for the TLS survey and around three weeks of full-time work to produce a DEM of $\approx$\num{80400}\si{m^2}.

SfM-MVS has gained attention recently for being a low-cost solution with fast and reliable results \citep{James2019}. We were able to cover $\approx$\num{740900}\si{m^2} with six RPA missions in under three hours. Processing time in a medium-range workstation (i.e., i7 processor, 64 GB RAM, dedicated GPU) was $\approx$13 hours. This makes it an excellent method for 3D modelling and continuous monitoring of coastal dunes. 

One strength of the ALS over TLS and SfM-MVS is the possibility of removing the vegetation based on the laser returns or waveform \citep{Brovelli2004,Evans2007,Khosravipour2016}, although new methods are being developed for single-return point clouds \citep{Guarnieri2009,Coveney2010,Coveney2011,Montreuil2013,Pijl2020} that have been used in coastal environments with good results \citep{Guisado-Pintado2019}. 

Another aspect to be considered is the weather. Dry and hot conditions will favour the presence of white sand patches, which can affect image matching and the 3D reconstruction.  While clear sunny days might be seen by many as ideal conditions for fieldwork, flying the RPA with cloudy skies and after a light rain can be worthwhile due the scattered light and visibility of the dune's superficial features.

\section*{Computer Code Availability}
Jupyter notebooks and associated data files (shapefiles and csv files) are available on GitHub (\url{https://github.com/CarlosGrohmann/scripts_papers/tree/master/garopaba_als_sfm_tls}) and Zenodo (\url{https://doi.org/10.5281/zenodo.3476779}). The notebooks are based on Python 3.6 and depend on the following libraries: numpy, scipy.stats, matplotlib, pandas, seaborn, rasterio, xarray, statsmodels, osgeo.ogr, pygrass and whiteboxtools.

\section*{Data Availability}
The point cloud datasets used in this study are available via the OpenTopography Facility\endnote{\url{http://opentopo.sdsc.edu}} \citep{Crosby2011,Krishnan2011}. The following datasets were used: OpenTopography ID OT.032013.32722.1 \citep[ALS --][]{Grohmann2010otopo,Grohmann2013gmorph}, OTDS.072019.32722.1 \citep[SfM --][]{Grohmann2019otopo-sfm-dune}, OTDS.102019.32722.1 \citep[TLS --][]{Grohmann2019otopo-tls-dune}.

% ----------------------------------------------------------------------------------
% ----------------------------------------------------------------------------------
% ----------------------------------------------------------------------------------
% ----------------------------------------------------------------------------------
\section*{Acknowledgements}

This study was supported by the Sao Paulo Research Foundation (FAPESP) grants \#2009/17675-5 and \#2016/06628-0 and by Brazil's National Council of Scientific and Technological Development, CNPq grants \#423481/2018-5 and \#304413/2018-6 to C.H.G. This study was financed in part by CAPES Brasil - Finance Code 001 through PhD scholarships to G.P.B.G, A.A.A. and R.W.A. This work acknowledges the services provided by the OpenTopography Facility with support from the National Science Foundation under NSF Award Numbers 1557484, 1557319, and 1557330. The authors are grateful to Clauson Lacerda and Rodrigo Correa (FARO Technologies Brasil), for their continuous support with terrestrial LiDAR data processing. John Lindsay (Un.Guelph) for discussions on FPD algorithm. Acknowledgements are extended to the Editor-in-Chief, the Associate Editor, and the anonymous reviewers for their criticism and suggestions, which helped to improve this paper.

% \section*{References}
% \bibliographystyle{elsarticle-harv}
% \bibliography{garopaba_sfm_v2r.bib}

% internet references
\theendnotes

\end{document}